\documentclass[twocolumn,showpacs,preprintnumbers,amsmath,amssymb]{revtex4}

\usepackage{graphicx}
\usepackage{dcolumn}
\usepackage{bm}

\newcommand{\bra}{\langle}
\newcommand{\ket}{\rangle}

\begin{document}


\title{Shear-induced quench of long-range correlations in a
 liquid mixture}

\author{Hirofumi Wada}

\affiliation{%
Department of Physics, University of Tokyo, Hongo, Tokyo, 113-0033,
Japan}%
\affiliation{Department of Mathematical and Life Sciences, 
Hiroshima University, Higashi Hiroshima, 739-8526, Japan%
}%
\date{\today}

\begin{abstract}
A static correlation function of concentration fluctuations
in a (dilute) binary liquid mixture subjected to both a concentration
gradient and uniform shear flow is investigated within the
framework of fluctuating hydrodynamics.
It is shown that a well-known $|\nabla c|^2/k^4$ 
long-range correlation at large wave numbers $k$ crosses over
to a weaker divergent one for wave numbers satisfying  
$k<(\dot{\gamma}/D)^{1/2}$, while an asymptotic shear-controlled 
power-law dependence is confirmed at much smaller wave numbers given 
by $k\ll (\dot{\gamma}/\nu)^{1/2}$,
where $c$, $\dot{\gamma}$, $D$ and $\nu$ are the mass concentration, 
the rate of the shear, the mass diffusivity and the kinematic viscosity 
of the mixture, respectively.
The result will provide for the first time the possibility
to observe the shear-induced suppression of a long-range correlation
experimentally by using, for example, a low-angle light
scattering technique. 
\end{abstract}

\maketitle

\section{\label{sec:level1} Introduction}

Long-range correlations in a fluid far from equilibrium
is one of the most salient features
of nonequilibrium fluctuations \cite{review1,review2,oppen,
ronis1,machta,siggia}.
Since the seminal discovery of the $k^{-4}$ enhancement of
the Rayleigh component of the structure factor in a fluid 
with an uniform temperature gradient by Kirkpatrick, Cohen and 
Dorfman \cite{kirkpatrick}, where $k$ is the wave number, 
algebraic decays of correlation functions have been
found theoretically in a wide class of nonequilibrium systems
\cite{ronis2,law,lutsko1,garcia,vailati1,garrido,yoshimura}.
The existence of such generic long-range correlations also 
have been confirmed in a series of the detailed light scattering 
experiments in pure liquids \cite{law2, segre1} and in binary 
mixtures \cite{segre2}.
Although the agreement between the theoretical studies and
the experimental ones are sufficiently quantitative, 
it have been limited to relatively high wave number regions.

On the other hand, large-scale, long-time behavior of 
such nonequilibrium fluctuations is attracting much more 
attentions in recent years \cite{note1}.
Ultra-low-angle light scattering experiment performed by
Vailati and Giglio \cite{vailati2} have revealed an 
impressive gravity-induced
quench of the $k^{-4}$ divergence of the Rayleigh line intensity
at very small $k$ range \cite{segre3} in a binary mixture 
subjected to an uniform concentration gradient driven by the 
large Soret effect.
So far as we know, this is the first experimental demonstration
that true asymptotic behavior of a long-range correlation at
large distances is qualitatively different from that predicted
by the linear response theory at short distances.

However, from the theoretical point of view, the gravity-induced
effect seems to be somewhat exceptional because it can be understood 
within the simple linear response theory.
Generally speaking, a full nonlinear analysis of hydrodynamic equations
is required for studying large-scale transport and fluctuation
properties of a fluid \cite{nelson}.
Therefore, mainly because of its technical difficulty, studies
beyond linear responses are rare in systems far from equilibrium
so far.
Only one exception is a system undergoing uniform shear flow where
effects of the shear advection can be treated in an
analytic way \cite{onuki1,onuki2,dufty}.
It has been predicted theoretically that the density
autocorrelation function and the density-momentum correlation
function in a sheared compressible fluid exhibit 
much weaker divergent behavior in small wave number limits than 
the $k^{-4}$ divergence found in large wave numbers \cite{lutsko2}.
A similar result has also been predicted for an incompressible fluid
\cite{wada}.
Nevertheless, it might be difficult to verify
these predictions directly by using available 
experimental techniques, because the crossover length scales
of these systems are so macroscopic.

The purpose of this paper is to emphasize effects of
the shear upon long-range correlations which can allow an
experimental confirmation.
To this end, we consider the particularly simple system,
a (dilute) low-molecular weight binary mixture with 
imposed concentration and uniform shear gradients 
\cite{note02}.
On the basis of fluctuating hydrodynamics approach \cite{landau}, 
we calculate an static concentration autocorrelation 
function which contributes to the total intensity of
the Rayleigh peak in a light scattering experiment.
In Sec. \ref{sec:model}, hydrodynamic equations for the 
concentration and the velocity field are introduced with 
suitable random forces.
We obtain general forms of space-time correlation functions
of the hydrodynamic variables by solving the linearized equations
around the steady state under imposed boundary conditions.
In Sec. \ref{sec:long}, the static correlation function of 
concentration fluctuations is expressed in terms of a universal
function which describes a shear-induced frustration of the 
long-range correlation.
This function is numerically evaluated and is compared with 
asymptotic functions that are analytically derived.
In Sec. \ref{sec:discussion}, we provide a simple physical 
interpretation of our results.
We also present rough estimations of the crossover length 
scales for a realistic fluid to suggest possible experimental 
verifications.
Our conclusion is given in the last section.

\section{\label{sec:model} Model and Analysis}

\subsection{Basic equations and approximations}

The mass diffusion flux ${\bf j}$ generally depends not only
on the concentration gradient $\nabla c$ but also on the 
temperature gradient $\nabla T$ and on the pressure gradient 
$\nabla p$ \cite{landau, mazur}.
However, when the mass concentration $c$ 
is small,
the thermal diffusion ratio is expected to be rather small
because it must vanish in a pure liquid 
\cite{landau}.
We thus can safely neglect the term proportional to $\nabla T$,
as well as the term proportional to $\nabla p$ in the expression
of ${\bf j}$ because we do not consider any gravity-induced
effect.
To make the analysis below as simple as possible, we 
further assume that the mixture is incompressible.
Consequently, the only relevant hydrodynamic variables are 
the concentration $c$ and the momentum $\rho{\bf v}$ of the mixture.
The equations describing the time evolution of these variables
supplemented by the suitable random forces are of the form
\cite{law,vailati1,brogioli}
\begin{eqnarray}
 \frac{\partial c}{\partial t}+{\bf v}\cdot\nabla c &=& 
	-\frac{1}{\rho}\nabla\cdot{\bf j}+\nabla\cdot{\bf g},
 \label{eq-c01}\\
 \frac{\partial{\bf v}}{\partial t}+({\bf v}\cdot\nabla){\bf v} 
 	&=& -\frac{1}{\rho}\nabla p+\nu\nabla^2{\bf v}+
		\nabla\cdot{\bf S},
 \label{eq-v01}
\end{eqnarray}
where the diffusion flux is given as 
${\bf j}\cong -\rho D\nabla c$,
and the hydrostatic pressure $p$ is determined from the
incompressibility condition $\nabla\cdot{\bf v}=0$.
Here $D$ is the mass diffusive coefficient, $\nu=\eta/\rho$
is the kinematic viscosity of the mixture ($\eta$ the zero 
shear viscosity).
The random forces ${\bf g}$ and ${\bf S}$ are the random 
concentration flux and the random stress tensor, respectively.
The correlations of these random forces retain their local
equilibrium values given by \cite{siggia,ronis2,law,landau}
\begin{eqnarray}
 \bra g_i({\bf r},t)g_j({\bf r}',t')\ket &=& 2k_BT\rho^{-1}
	D\chi_c\delta_{ij} \nonumber \\
 & & \times \delta^3({\bf r}-{\bf r}')
		\delta(t-t'),
 \label{fdt-c01}\\
 \bra S_{il}({\bf r},t)S_{jm}({\bf r}',t')\ket &=& 2k_BT
	\rho^{-1}\nu\left[\delta_{ij}\delta_{lm}+\delta_{im}
		\delta_{jl}\right. \nonumber \\
 & & \!\!\!\!\!\! \left.-{\textstyle{\frac{2}{3}}}
			\delta_{il}\delta_{jm}\right]
				\delta^3({\bf r}-{\bf r}')
					\delta(t-t'),
 \label{fdt-v01}
\end{eqnarray}
and ${\bf g}$ is uncorrelated with ${\bf S}$.
In the following, we assume that the osmotic compressibility
$\chi_c=(\partial c/\partial \mu)_{p,T}$ and the mass diffusivity
$D$ are both independent of 
the concentration $c$, where $\mu$ is the chemical 
potential of the mixture \cite{landau,law,vailati1}.
This assumption may be valid when the gradient $|\nabla c|$
is not sufficiently large.
We note that the local equilibrium assumption in
Eqs. (\ref{fdt-c01}) and (\ref{fdt-v01}) is justified from
the fact that the random forces represent the
fast, and localized molecular process which cannot be affected
by the macroscopic gradients of the hydrodynamic quantities.

\subsection{Boundary conditions}

Although we are primarily interested in the bulk fluctuation
properties of the mixture, appropriate boundary conditions 
are required to specify a macroscopic steady state of
the averaged thermodynamic quantities \cite{siggia}.
We adopt here the standard Lees-Edwards type
boundary conditions \cite{lutsko2,edwards}, which can be
written explicitly as
\begin{equation}
 {\bf v}(x,y={\textstyle{\frac{1}{2}L}},z) =
	{\bf v}(x-\dot{\gamma}Lt,y=-{\textstyle{\frac{1}{2}L}},z)+
		\dot{\gamma}L\hat{{\bf e}}_x,
 \label{bc-v}
\end{equation}
and
\begin{equation}
 c(x,y={\textstyle{\frac{1}{2}L}},z) =
	c(x-\dot{\gamma}Lt,y=-{\textstyle{\frac{1}{2}L}},z)+
		|\nabla c|L,
 \label{bc-c}
\end{equation}
where $L$ is the linear system size of $y$ direction that
is parallel to both the concentration and shear gradient 
directions.
It is also assumed that the system is infinitely large
in all other directions.
The statistically averaged, time-independent equations 
(\ref{eq-c01}) and (\ref{eq-v01}) under the imposed 
boundary conditions (\ref{bc-v}) and (\ref{bc-c}) have 
solutions
\begin{eqnarray}
 \bra{\bf v}({\bf r})\ket &=& \dot{\gamma}y\hat{{\bf e}}_x,
 \label{avsol-v}\\
 \bra{c}({\bf r})\ket &=& c_0+|\nabla c|y,
 \label{avsol-c}
\end{eqnarray}
where $c_0$ stands for the concentration in the $y=0$ plane.
We here chosen our coordinate frame whose origin is at
rest by making use of the Galilean invariance. 
It is also easy to show that, within the present model, the 
average concentration profile (\ref{avsol-c}) is linearly
stable for a perturbation with any wave number under
the imposed velocity profile (\ref{avsol-v}) \cite{note3}.

\subsection{Linearized hydrodynamic equations around the
steady state}
Now let $c=\bra{c}\ket+\delta c$ and 
${\bf v}=\bra{\bf v}\ket+\delta{\bf v}$ 
in Eqs. (\ref{eq-c01}) and (\ref{eq-v01}).
We then obtain the linearized equations for fluctuations
$\delta c$ and $\delta {\bf v}=(\delta u,\delta v,\delta w)$ 
as
\begin{eqnarray}
 \frac{\partial}{\partial t}\delta c+\dot{\gamma}y
	\frac{\partial}{\partial x}\delta c &=&
		-|\nabla c|\delta v+D\nabla^2\delta c+\theta,
 \label{eq-c02}\\
 \frac{\partial}{\partial t}\delta{\bf v}+\dot{\gamma}y
	\frac{\partial}{\partial x}\delta{\bf v}
		+\dot{\gamma}\delta v\hat{{\bf e}}_x &=&
		-\frac{1}{\rho}\nabla\delta p+\nu\nabla^2
			\delta{\bf v}+{\bf f},
 \label{eq-v02}
\end{eqnarray}
where an equation for the pressure fluctuation follows from
the divergence of the equation for $\delta{\bf v}$:
\begin{equation}
 \nabla^2\delta p = -2\rho\dot{\gamma}\frac{\partial}{\partial x}
	\delta v+\rho\nabla\cdot{\bf f}.
 \label{press}
\end{equation}
Here we introduced the random variables $\theta=\nabla\cdot
{\bf g}$ and ${\bf f}=\nabla\cdot{\bf S}$.
Equation (\ref{eq-c02}) implies that the only $y$
component of the velocity fluctuation $\delta v$ 
can affect the dynamics of $\delta c$.
Eliminating $\delta p$ from Eq. (\ref{eq-v02}) by 
using Eq. (\ref{press}), we obtain the linearized hydrodynamic 
equations relevant to the present purpose of the form
\begin{eqnarray}
 \left(\frac{\partial}{\partial t}-\dot{\gamma}
	k_x\frac{\partial}{\partial k_y}\right)\delta
		c_{\bf k}(t) &=& -|\nabla c|\delta
			v_{\bf k}(t) \nonumber \\
 & & -Dk^2\delta c_{\bf k}(t)+\theta_{\bf k}(t),
 \label{eq-c03}\\
 \left(\frac{\partial}{\partial t}-\dot{\gamma}
	k_x\frac{\partial}{\partial k_y}\right)\delta
		v_{\bf k}(t) &=& -\left(\nu k^2-2\dot{\gamma}
			\frac{k_xk_y}{k^2}\right)
				\delta v_{\bf k}(t) \nonumber \\
 & & +f_{\bf k}(t),
 \label{eq-v03}
\end{eqnarray}
where the Fourier transform of an arbitrary function
$\phi({\bf r},t)$ is defined as
\begin{equation}
 \phi({\bf r},t) = \int\frac{d{\bf k}}{(2\pi)^3}
	\phi_{\bf k}(t)e^{i{\bf k}\cdot{\bf r}}.
 \label{fourier}
\end{equation}
Note that the random variable in Eq. (\ref{eq-v03}) is
re-defined by $f_{\bf k}(t)=\sum_j(\delta_{yj}-k_yk_j/k^2)
f_{j{\bf k}}(t)$.
This variable and $\theta_{\bf k}(t)$ satisfy the 
correlation properties given from Eq. (\ref{fdt-c01}) 
and Eq. (\ref{fdt-v01}) as
\begin{eqnarray}
 \bra\theta_{\bf k}(t)\theta_{{\bf k}'}(t')\ket &=&
	2k_BT\rho^{-1}D\chi_ck^2 \nonumber\\
 &\times& (2\pi)^3\delta^3({\bf k}+{\bf k}')\delta(t-t'),
 \label{fdt-c02}\\
 \bra f_{\bf k}(t)f_{{\bf k}'}(t')\ket &=&
	2k_BT\rho^{-1}\nu k^2_{\perp} \nonumber \\
 &\times& (2\pi)^3\delta^3({\bf k}+{\bf k}')\delta(t-t'),
 \label{fdt-v02}
\end{eqnarray}
and $\bra \theta_{{\bf k}}(t)f_{{\bf k}'}(t')\ket = 0$,
where $k_{\perp}^2=k_x^2+k_z^2$.

The set of equations (\ref{eq-c03}) and (\ref{eq-v03})
are most easily solved by making a transformation to the
local Lagrangian coordinate given by 
${\bf r}'={\bf r}-\dot{\gamma}yt\hat{\bf e}_x$ 
\cite{lutsko1,onuki1,onuki2,dufty}.
Applying the standard manipulation described elsewhere
\cite{dufty}, we readily find the solutions of the form
\begin{eqnarray}
 \delta v_{\bf k}(t) &=& \int_0^{\infty} 
	f_{{\bf k}(-s)}(t-s)G_{vv}({\bf k},s)ds,
 \label{sol-v01}\\
 \delta c_{\bf k}(t) &=& \delta c^{(0)}_{\bf k}(t)-|\nabla c|
	\delta c^{(1)}_{\bf k}(t),
 \label{sol-c01}
\end{eqnarray}
where 
\begin{equation}
 \delta c^{(0)}_{\bf k}(t) = \int_0^{\infty}
	\theta_{{\bf k}(-s)}(t-s)G_{cc}({\bf k},s)ds,
 \label{sol-c01a}\\
\end{equation}
and 
\begin{equation}
 \delta c^{(1)}_{\bf k}(t) = \int_0^{\infty}
	\delta v_{{\bf k}(-s)}(t-s)G_{cc}({\bf k},s)ds.
 \label{sol-c01b}
\end{equation}
Here ${\bf k}(t)$ is the time-dependent wave vector
defined as
\begin{equation}
 {\bf k}(t) = {\bf k}-\dot{\gamma}k_xt\hat{\bf e}_y,
 \label{kt}
\end{equation}
and the Green's functions are given by
\begin{equation}
 G_{\alpha\alpha}({\bf k},t) = \exp\left(-\int_0^t
	\Gamma_{\alpha\alpha}({\bf k}(-\tau))d\tau\right),
 \label{green}
\end{equation}
where $\Gamma_{cc}(k)=Dk^2$ and  
$\Gamma_{vv}(k)=\nu k^2-2\dot{\gamma}k^{-2}k_xk_y$.

\subsection{Correlation functions}
The space-time correlation function of the hydrodynamic
fluctuations under the uniform shear is defined by \cite{lutsko1}
\begin{equation}
 \tilde{C}_{\alpha\beta}({\bf k},t;{\bf k}',t')=
	\bra\phi_{\alpha,{\bf k}}(t)\phi_{\beta,{\bf k}'}
		(t')\ket
 \label{cf01}
\end{equation}
for $(\alpha,\beta)=(c,v)$, where the angle bracket represents
the statistical average with respect to the random variables
$\theta$ and $f$.
The function $\phi_{\alpha}$ represents either
$\delta c$ or $\delta v$.
Under the uniform shear condition, the system is invariant
with respect to the temporal transformation, whereas
it is not invariant with respect to the spatial 
transformations \cite{lutsko1,onuki2,dufty,lutsko2}.
The time invariance property imposes the symmetry
\begin{equation}
 \tilde{C}_{\alpha\beta}({\bf k},t;{\bf k}',t')=
	\tilde{C}_{\alpha\beta}({\bf k},t-t';{\bf k}',0).
 \label{symmet}
\end{equation}
This relation implies that it is sufficient to 
consider the function 
$\tilde{C}_{\alpha\beta}({\bf k},t;{\bf k}',0)\equiv
\tilde{C}_{\alpha\beta}({\bf k},{\bf k}',t)$ for $t\geq0$
in the following calculations, without loss of 
generality \cite{lutsko1}.

Substituting Eqs. (\ref{sol-v01}) and (\ref{sol-c01}) 
into Eqs. (\ref{cf01}) and using the properties of the 
random variables (\ref{fdt-c02}) and (\ref{fdt-v02}), 
we can derive the time correlation functions of the
hydrodynamic modes.
The calculations are somewhat lengthy, so we here give
only the results:
\begin{equation}
 \tilde{C}_{\alpha\beta}({\bf k},{\bf k}',t) = 
	C_{\alpha\beta}({\bf k},t)
	(2\pi)^3\delta^3({\bf k}+{\bf k}'(t)),
 \label{cf02}
\end{equation}
where
\begin{widetext}
\begin{equation}
 C_{vv}({\bf k},t)=\frac{k_BT}{\rho}\frac{k^2_{\perp}}
	{k^2}\left[G_{vv}({\bf k},t)+
		2\dot{\gamma}G^{-1}_{vv}({\bf k},t)\int_t^{\infty}
			(\hat{k}_x\hat{k}_y+\hat{k}_x^2s)
				G^2_{vv}({\bf k},s)ds\right],
 \label{cf-v02}
\end{equation}
\begin{equation}
 C_{cc}({\bf k},t) = 
	\frac{k_BT}{\rho}\chi_c\left[G_{cc}({\bf k},t)
 		+2|\nabla c|^2\frac{(\nu k_{\perp}^2)}{\chi_c}
			G_{cc}^{-1}({\bf k},t)
	\int_0^{\infty}ds\frac{G_{cc}({\bf k},s)}
		{G_{vv}({\bf k},s)}\int_{\max{(s,t)}}
			^{\infty}ds'G^2_{vv}({\bf k},s') 
 \int_t^{s'}ds''\frac{G_{cc}({\bf k},s'')}
	{G_{vv}({\bf k},s'')}\right],
 \label{cf-c02}
\end{equation}
and the cross-correlation function given by
\begin{equation}
C_{cv}({\bf k},t)=-2\frac{k_BT}{\rho}|\nabla c|
	(\nu k^2_{\perp})G_{vv}^{-1}({\bf k},t)\int_0^{\infty}ds
		\frac{G_{cc}({\bf k},s)}{G_{vv}({\bf k},s)}
			\int_{\max{(s,t)}}^{\infty}ds'
				G^2_{vv}({\bf k},s').
 \label{cf-cv02}
\end{equation}
\end{widetext}
Here the Green's functions are written explicity as
\begin{eqnarray}
 G_{cc}({\bf k},t) &=& e^{-D k^2T(\hat{\bf k},t)},
 \label{green-c}\\
 G_{vv}({\bf k},t) &=& \tau(\hat{\bf k},t)
	e^{-\nu k^2T(\hat{\bf k},t)},
 \label{green-v}
\end{eqnarray}
with
\begin{eqnarray}
 T(\hat{\bf k},t) &=& t+\dot{\gamma}\hat{k}_x\hat{k}_yt^2
	+\frac{1}{3}{\dot{\gamma}}^2\hat{k}_x^2t^3,
 \label{T}\\
 \tau(\hat{\bf k},t) &=& \frac{\partial}{\partial t}
	T(\hat{\bf k},t) = 1+2\dot{\gamma}\hat{k}_x\hat{k}_yt
		+{\dot{\gamma}}^2\hat{k}_x^2t^2,
 \label{tau}
\end{eqnarray}
where $\hat{\bf k}={\bf k}/|{\bf k}|$.
In the derivations of Eqs. (\ref{cf-v02})-(\ref{cf-cv02}),
we have made use of the identity 
$G_{\alpha\alpha}({\bf k}(-t),s)=
G^{-1}_{\alpha\alpha}({\bf k},t)G_{\alpha\alpha}({\bf k},s+t)$.

\begin{figure}
\includegraphics[width=8cm]{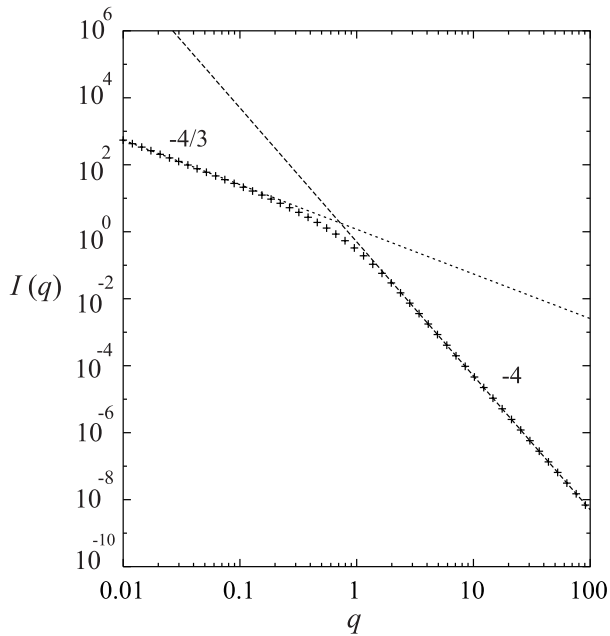}
\caption{\label{fig:v-v corr} Plots of numerically
evaluated scaling function $I(q)$, where the
momentum autocorrelation function is written in the 
form $C_{vv}(k,0,0)=\rho^{-1}k_BT[1+I(\xi_v k)]$.
The dashed and dotted lines represent the aymptotic 
functions for large $q\gg1$ (Eq. (\ref{cf-v05a})) and 
for small $q\ll1$ (Eq. (\ref{cf-v05b})), respectively.}
\end{figure}

Although Eqs. (\ref{cf-v02})-(\ref{cf-cv02}) are rather 
complicated, the equal-time correlation functions have 
somewhat simpler forms because of the recovery of the 
translational invariance with respect to the space.
Setting $t=0$ in Eq. (\ref{cf-v02})-(\ref{cf-cv02}), we obtain
\begin{widetext}
\begin{equation}
 C_{vv}({\bf k})=\frac{k_BT}{\rho}\frac{k^2_{\perp}}
	{k^2}\left[1+2\dot{\gamma}\int_0^{\infty}dt
			(\hat{k}_x\hat{k}_y+\hat{k}_x^2t)
				e^{-2\nu k^2T(\hat{\bf k},t)}\right],
 \label{cf-v03}
\end{equation}
\begin{equation}
 C_{cc}({\bf k}) =
	\frac{k_BT}{\rho}\chi_c\left[1
 		+2|\nabla c|^2\frac{(\nu k_{\perp}^2)}{\chi_c}
			\int_0^{\infty}dt
 \frac{e^{(\nu-D)k^2T(\hat{\bf k},t)}}{\tau(\hat{{\bf k}},t)}
	\int_t^{\infty}dt'\tau^2(\hat{{\bf k}},t')
		e^{-2\nu k^2T(\hat{\bf k},t')}\int_0^{t'}dt''
	\frac{e^{(\nu-D)k^2T(\hat{\bf k},t'')}}
		{\tau(\hat{{\bf k}},t'')}\right],
 \label{cf-c03}
\end{equation}
and
\begin{equation}
 C_{cv}({\bf k})=-2\frac{k_BT}{\rho}|\nabla c|
	(\nu k^2_{\perp})\int_0^{\infty}dt
		\frac{e^{(\nu-D)k^2T(\hat{\bf k},t)}}
			{\tau(\hat{{\bf k}},t)}
				\int_t^{\infty}dt'
					\tau^2(\hat{{\bf k}},t')
 e^{-2\nu k^2T(\hat{\bf k},t')}.
 \label{cf-cv03}
\end{equation}
Equations (\ref{cf-v03})-(\ref{cf-cv03}) together with
Eqs. (\ref{T}) and (\ref{tau}) are the main results of this
section.
The same result as Eq. (\ref{cf-v03}) has also been derived
for the simple fluid in Ref. \cite{wada}.
Thus the correlation function between the momentum fluctuations
is unaffected by the steady concentration gradient within 
the present analysis.

\section{\label{sec:long} long-range correlations}
We shall first examine the two limiting cases studied
previously in Eqs. (\ref{cf-v03})-(\ref{cf-cv03}).
At zero shear rate, 
Eqs. (\ref{cf-v03})-(\ref{cf-cv03}) are reduced to the form
\begin{eqnarray}
 C_{vv}({\bf k}) &\rightarrow& \frac{k_BT}{\rho}
	|\hat{{\bf k}}_{\perp}|^2,
 \label{cf-v04}\\
 C_{cc}({\bf k}) &\rightarrow& \frac{k_BT}{\rho}\chi_c\left[1
 		+2|\nabla c|^2\frac{(\nu k_{\perp}^2)}{\chi_c}
			\int_0^{\infty}dt
 e^{(\nu-D)k^2t}\int_t^{\infty}dt'e^{-2\nu k^2t'}\int_0^{t'}
	dt''e^{(\nu-D)k^2t''}\right]\nonumber\\
 &=& \frac{k_BT}{\rho}\chi_c\left(1+\frac{|\nabla c|^2}
	{\chi_cD(\nu+D)}\frac{|\hat{{\bf k}}_{\perp}|^2}{k^4}\right),
 \label{cf-c04}\\
 C_{cv}({\bf k}) &\rightarrow& -2\frac{k_BT}{\rho}
	|\nabla c|(\nu k_{\perp}^2)\int_0^{\infty}dt
 		e^{(\nu-D)k^2t}\int_t^{\infty}dt'
			e^{-2\nu k^2t'}
 = -\frac{k_BT}{\rho}\frac{|\nabla c|}{(\nu+D)}
	\frac{|\hat{{\bf k}}_{\perp}|^2}{k^2}.
 \label{cf-cv04}
\end{eqnarray}
\end{widetext}
Equation (\ref{cf-v04}) is the standard result in thermal
equilibrium \cite{forster}.
In contrast to this, the second term in Eq. (\ref{cf-c04})
and Eq. (\ref{cf-cv04}) show the anomalous enhancements of 
the hydrodynamic fluctuations in the presence of the 
concentration gradient.
These results are first predicted by Kirkpatrick {\it et al}. 
using mode coupling and kinetic theory \cite{kirkpatrick} 
and subsequently confirmed by Ronis and Procaccia using fluctuating 
hydrodynamics \cite{ronis2}.
On the other hand, in the limit of a small concentration gradient, 
Eqs. (\ref{cf-c04}) and (\ref{cf-cv04}) converge to the 
equilibrium forms given by
$C_{cc}({\bf k}) \rightarrow k_BT\chi_c/\rho$ and 
$C_{cv}({\bf k}) \rightarrow 0$ regardless of the presence
of the shear flow.
However, the momentum-momentum correlation function
becomes long-ranged in this case because of the influence
of the shear \cite{siggia,lutsko1,lutsko2,wada,ronis2}.
Two asymptotic forms in the selected direction of the
wave vector ${\bf k}=(k,0,0)$ are given by \cite{wada} 
\begin{equation}
 C_{vv}(k,0,0) \sim \frac{k_BT}{\rho}\left(1+\frac{1}{2}
	\frac{{\dot{\gamma}^2}}{\nu^2k^4}\right)
 \label{cf-v05a}
\end{equation}
for $k\xi_v \gg 1$, and
\begin{equation}
 C_{vv}(k,0,0) \sim \frac{k_BT}{\rho}\left(\frac{2}{3}
	\right)^{1/3}\Gamma\left(\frac{2}{3}\right)
	\frac{{\dot{\gamma}^{2/3}}}{\nu^{2/3}k^{4/3}}
 \label{cf-v05b}
\end{equation} 
for $k\xi_v \ll 1$, where $\Gamma(x)$ is the Gamma function.
The direct numerical evaluation of Eq. (\ref{cf-v03}) is also 
shown in Fig. \ref{fig:v-v corr}.
The length scale which characterize this crossover is
introduced as
\begin{equation}
 \xi_v = \sqrt{\frac{\dot{\gamma}}{\nu}}.
 \label{xi-v}
\end{equation}

In the following, we shall focus our attention to the behavior of 
Eq. (\ref{cf-c03}) for arbitrary magnitudes of the shear rate 
$\dot{\gamma}$ and the concentration gradient $|\nabla c|$.
It is easy to show that Eq. (\ref{cf-c03}) can be
written in the form
\begin{equation}
 C_{cc}({\bf k}) = \frac{k_BT}{\rho}\chi_c\left(1+
	\frac{|\nabla c|^2}{\chi_c{\dot{\gamma}}^2}
		F({\bf k}\xi_v)\right).
 \label{cf-c05}
\end{equation}
We here consider the case where the
scattering wave vector ${\bf k}$ is perpendicular to the
concentration gradient $\nabla c$, i.e., 
${\bf k}={\bf k}_{\perp}=(k_x,0,k_z)$.
This scattering geometry still holds all of the essential 
features of the nonequilibrium effects found in this
system, and is actually the configuration adopted
in most of the previous light scattering experiments 
\cite{law,law2,segre1,segre2,vailati2}.
The scaling function $F({\bf q})$ in Eq. (\ref{cf-c05}) 
in this geometry is given by
\begin{widetext}
\begin{equation}
 F({\bf q}) = 2q^2\int_0^{\infty}\frac{ds}{1+\hat{q}_x^2
	s^2}e^{(1-a)q^2\tilde{T}(\hat{\bf q},s)}
		\int_s^{\infty}ds'\left(1+
			\hat{q}_x^2{s'}^2\right)^2e^{-2q^2
				\tilde{T}(\hat{\bf q},s')} 
 \int_0^{s'}\frac{ds''}{1+\hat{q}_x^2
	{s''}^2}e^{(1-a)q^2\tilde{T}(\hat{\bf q},s'')},
 \label{F01}
\end{equation}
\end{widetext}
where $\tilde{T}(\hat{\bf q},t)=t+\frac{1}{3}\hat{q}_x^2t^3$
and $a=D/\nu$ is the ratio of the mass diffusivity to the
viscous diffusivity, which is usually much smaller than the
unity.
As can be understood from Eq. (\ref{F01}), the effect of the
shear is most likely to become evident in the particular
direction of the wave vector given by ${\bf k}=(k,0,0)$.
To simplify the analysis, we restrict our interest to the 
fluctuations in this direction.
Because of this simplification, the asymptotic behavior 
for small $q=k\xi_v$ limit is easily found to be 
\begin{equation}
 F(q) \sim \left(\frac{3}{2}\right)^{2/3}\frac{\pi^2}{4}
	\Gamma\left(\frac{5}{3}\right)q^{-4/3}.
 \label{F-as01}
\end{equation}
Equation (\ref{F-as01}) suggests that the enhancement of
the nonequilibrium fluctuations is severely restricted
in the long wavelength limit.
On the other hand, the opposite asymptotic region, i.e.,
the short wavelength limit, is represented by $k\xi_c \gg 1$,
where the second characteristic length scale $\xi_c$ is
given by
\begin{equation}
 \xi_c = \sqrt{\frac{D}{\dot{\gamma}}} = a^{1/2}\xi_v
	\ll \xi_v.
 \label{xi_c}
\end{equation} 
It turns out that the asymptotic behavior
of $F(k\xi_v)$ for $k\xi_c \gg 1$ has the same 
wave number dependence given by  
the second term in Eq. (\ref{cf-c04}).
This implies that thermal fluctuations of the hydrodynamic 
variables for $k\xi_c \gg 1$ are little affected by the
shear and can dissipate purely thermally.  
Consequently, the static concentration autocorrelation function 
in the two limiting cases of short and long wave numbers
are written as
\begin{equation}
 C_{cc}(k,0,0) \sim \frac{k_BT}{\rho}\chi_c\left(
	1+\frac{|\nabla c|^2}{\chi_cD(\nu+D)k^4}\right)
 \label{cf-c06a}
\end{equation}
for $k\xi_v \gg 1$, and
\begin{equation}
 C_{cc}(k,0,0) \sim \frac{k_BT}{\rho}\chi_c\left(
	1+\frac{\alpha|\nabla c|^2}{\chi_c\nu^{2/3}
		{\dot{\gamma}}^{4/3}k^{4/3}}\right)
  \label{cf-c06b}
\end{equation}
for $k\xi_c \ll 1$,
where the numerical constant $\alpha$ is given by
$\alpha=(\frac{3}{2})^{2/3}\frac{\pi^2}{4}\Gamma
(\frac{5}{3}) \approx 2.9$.
\begin{center}
\begin{figure}
\includegraphics[width=8.5cm]{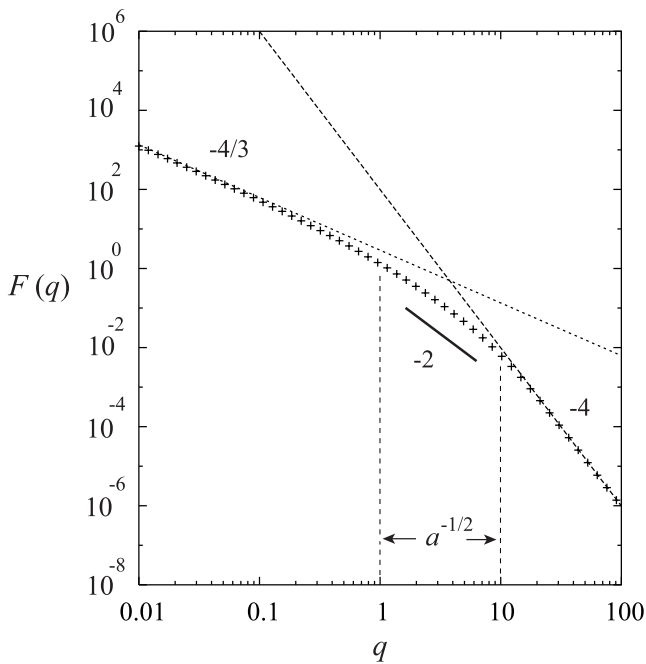}
\caption{\label{fig:c-c corr} The scaling function
$F(k\xi_v)$ as a function of $q=k\xi_v$ obtained by the
numerical integration of Eq. (\ref{F01}) for $\hat{q_x}=1$ 
and $a=0.01$. The lines are the asymptotic results derived
analytically.
Note that the crossover wave numbers given by $k\xi_v=1$
and $k\xi_c=1$ correspond to $q=1$ and $q=a^{-1/2}$,
respectively.}
\end{figure}
\end{center}
In Fig. \ref{fig:c-c corr}, a numerically calculated scaling function
$F(q)$ is plotted with the asymptotic functions
for large and small $q$ deduced from Eqs. (\ref{cf-c06a})
and (\ref{cf-c06b}), respectively.
Although the intermediate region given by 
$\xi_v^{-1}<k< \xi_c^{-1}$ seems to have 
$k^{-2}$ dependence, whether it is a true scaling regime 
or just a crossover is still unclear at present.
Note, however, that this region is practically extended
in a wide range of the wave numbers, which is typically 
given as $1<q<a^{-1/2}(\sim 10^2-10^3)$ due to the large 
asymmetry between the magnitude of the viscous and 
mass diffusivity.
Therefore, we expect that this unclearness will be clarified
by evaluating $F(q)$ for a much smaller $a$ than that considered
here ($a=10^{-2}$), though it cannot be done
in the present study primarily because of the limitation of 
the numerical accuracy. 
 
\section{\label{sec:discussion} discussion}
As pointed out in a number of literature, the long-range
correlation is 
originated from the coupling of momentum fluctuations with 
concentration fluctuations through a macroscopic 
concentration gradient \cite{kirkpatrick,ronis2}.
This is sometimes called a linear mode-coupling effect
\cite{lutsko1}.
Following the discussions given in Ref. \cite{weitz,brogioli2},
let us suppose a small fluid element of linear size 
$\xi$ in the mixture in the absence of the shear.
This fluid portion is carried along the 
direction of the gradient by a spontaneously generated 
momentum fluctuation of its lifetime given by 
$\tau_v\sim \xi^2/\nu$.
Because the surrounding fluid has a different
average concentration, the large concentration difference
persists for a time $\tau_c \sim \xi^2/D$ as
it relaxes purely diffusively.
Considering that $a=D/\nu \ll 1$, one can notice that
a short-living, spontaneous momentum fluctuation can 
creates a long-lasting, large-amplitude concentration 
fluctuation, which results in a long-range spatial 
correlation between concentration fluctuations 
in the steady state.

However, shear flow limits the size of such fluctuations
significantly,
because a sufficiently large fluctuation is drawn out 
and is even broken up by the shear before it disappears
by the diffusion \cite{onuki1,onuki2,lutsko2}.
In our system, $\xi_c$ and $\xi_v$ correspond to the 
crossover length scale from the diffusion-dominated decay
to the shear-dominated decay of concentration and momentum
fluctuations, respectively. 
Because $\xi_c$ is much smaller than $\xi_v$ in practice,
the overall behavior of $C_{cc}({\bf k})$ is expected to
have three distinctive wave number 
domains. They can be sketched as follows.

(I) When a small portion of mixture of linear size larger 
than $\xi_v$ is carried along the direction of the gradient, 
it undergoes a significant shear deformation and becomes 
highly anisotropic.
Because the lifetime of the concentration fluctuation is 
controlled by the shortest length scale of its spatial 
extent, this fluctuation 
dissipates thermally much faster than that in a quiescent
fluid.
Thus the long-range correlation between concentration
fluctuations is severely suppressed by the shear flow
in this wave number region. 

(II) When $\dot{\gamma}\tau_v\sim \xi^2/\xi_v^2<1$,
a momentum fluctuation can displace a small parcel of
fluid of size $\xi$ without a notable shear deformation.
However, if $\dot{\gamma}\tau_c\sim \xi^2/\xi_c^2>1$,
the shear flow strongly affects the decay of this
large-amplitude concentration fluctuation before it
dissipates thermally.
Then the decay of the fluctuation is still faster than
that in a quiescent fluid in this region, and the 
resulting correlation exhibits a weaker divergent 
dependence on the wave numbers $k$ than $k^{-4}$.

(III) When a size of a fluctuation is much smaller than 
$\xi_c$, the effect of the shear becomes rather weak 
because a thermal diffusive decay is faster than the 
shear deformation time scale; 
$\dot{\gamma}\tau_v\ll\dot{\gamma}\tau_c<1$. 
Therefore the usual linear response argument 
for a quiescent fluid presented in the beginning of this section 
is fit for this regime.

Figure \ref{fig:c-c corr} shows that the deviation from the $k^{-4}$ 
divergence becomes pronounced at the wave number $k_c \sim \xi_c^{-1}$.
It should be emphasized that the presence of $k_c$ is
the clear evidence of the shear-induced suppression of 
the long-range correlation.
Although the shear-controlled asymptote is confirmed
only in the wave numbers smaller than $k_v\sim\xi_v^{-1}$, it 
may be impossible to verify this scaling experimentally
because the length scale $\xi_v$ is so macroscopic for 
relevant fluid parameters and experimentally feasible shear 
rates \cite{lutsko2,wada}.
(For example, we can obtain $k_v \sim 1$ mm$^{-1}$ at most, for
a typical value of the viscosity $\nu \sim 10^{-2}$ cm$^2$s$^{-1}$ 
and for a very high rate of the shear that is of the 
order of $10^4$ s$^{-1}$.)
Contrary to this situation, a typical value of the 
diffusion coefficient $D \sim 10^{-6}$ cm$^2$s$^{-1}$
gives $k_c\sim 10^3$ cm$^{-1}$ even for a very small
rate of the shear as $\dot{\gamma}\sim 1$ s$^{-1}$.
Therefore, the wave number $k_c$, which characterizes 
the onset of the shear-induced frustration of the 
long-range correlation, is well covered by conventional
low-angle light scattering methods, as well as the 
recently reported new optical technique \cite{brogioli3}.

There have been developed some experimental techniques to
produce a macroscopic concentration gradient
in a fluid, such as utilizing a diffusive remixing process
of a mixture \cite{brogioli2} or making use of a large Soret 
effect driven by a steady temperature 
gradient \cite{vailati2,cerbino}.
Although it is far beyond our ability to guess about, the latter
method might be more favorable with respect to the 
compatibility with the shear flow, as well as the spatial
uniformity and the temporal stationarity of the concentration
gradient.
When the large Soret effect is expected, however, we have to 
take a temperature fluctuation into account as well, 
because the cross-correlation between
concentration and temperature fluctuation makes an important
contribution to the Rayleigh scattering \cite{law,segre2}.
In that case, the third length scale 
$\xi_T=(D_T/\dot{\gamma})^{1/2}$ may enter into the 
theory, where $D_T$ is the thermal diffusivity of the mixture.
Nevertheless, since $\nu \ll D_T \ll D$ for most of 
relevant mixtures, 
$\xi_c$ is likely to be well separated from the other two 
length scales. 
Then, so far as we are concerned with the 
shear-induced effect, especially with the onset of the 
crossover, we can expect that the present analysis is 
applicable to such cases.

\section{conclusion}
In this paper, using fluctuating hydrodynamics, 
we have studied the fluctuation properties in a (dilute)
binary mixture with imposed uniform concentration and shear
gradients.
It is demonstrated that the static concentration autocorrelation 
function has the three distinctive wave number regions.
The length scales which characterize the crossovers 
between different regimes are given as $\xi_c\sim 
(D/\dot{\gamma})^{1/2}$ and $\xi_v\sim 
(\nu/\dot{\gamma})^{1/2}$, respectively.
In particular, shear strongly suppresses the $k^{-4}$
divergence for $k\xi_c <1$,
while the asymptotic shear-controlled behavior is found
analytically in the smallest wave numbers satisfying
 $k\xi_v \ll 1$.
It is worth noting that the onset of
the shear-induced quench of the long-range correlation
may be experimentally observable.
Such experimental verification will also become an important
test for the applicability of fluctuating hydrodynamics to 
a fluid in a coupled-nonequilibrium steady state.

\begin{acknowledgments}
The author is grateful to S-i. Sasa for guiding him in the
problem of nonequilibrium fluctuations, and a number of
valuable discussions.
The author also acknowledges R. Cerbino and A. Vailati
for introducing their original works and illuminating
discussions which gave a motive for the present study
at the visit in University of Milan.
Thanks are due to T. Ohta and T. Shibata for their
hospitality in Hiroshima University, where much of this
work has been carried out.
Lastly, the author would like to thank M. Sano for his 
encouragements.
This work was supported by Grant-in-Aid for JSPS
Fellows for Young Scientists, from Ministry of Education,
Culture, Sports, Science, and Technology, Japan.
\end{acknowledgments}


\begin{thebibliography}{99}
\bibitem{note1} It is worth noting that such a 
motivation comes in part from the insight on the possible
relationship between a non-analytic dependence of a 
thermodynamic quantity on a nonequilibrium control parameter
and a long-range correlation of its fluctuation.
See also Ref. \cite{wada}.
\end{thebibliography}

\end{document}